\documentclass[a4paper]{spie}
\usepackage[dvips]{graphicx}
\usepackage{epsfig}

\title{Mid-Infrared Instrumentation for the European Extremely Large Telescope}
\author{S. Kendrew\supit{a}, B. Brandl\supit{a}, R. Lenzen\supit{b}, L. Venema\supit{c}, H.~U. K\"{a}ufl\supit{d}, G. Finger\supit{d}, A. Glasse\supit{e}, R. Stuik\supit{a}}

\begin{document}
\maketitle

\begin{center}
\supit{a}Leiden Observatory, University of Leiden, PO Box 9513, 2300 RA Leiden, Netherlands\\
\supit{b}Max Planck Institute for Astronomy, Koenigstuhl 17, 69117 Heidelberg, Germany\\
\supit{c}Astron, PO Box 2, 7990 AA Dwingeloo, Netherlands\\
\supit{d}ESO, Karl Schwarzschildstrasse 2, 85748 Garching, Germany\\
\supit{e}UK Astronomy Technology Centre, Blackford Hill, Edinburgh EH9 3HJ, United Kingdom
\end{center}
\authorinfo{Send correspondence to: S. Kendrew, Leiden Observatory, kendrew@strw.leidenuniv.nl, tel: +31-71-5278456}

\abstract{MIDIR is the proposed thermal/mid-IR imager and spectrograph for the European Extremely Large Telescope (E-ELT). It will cover the wavelength range of 3 to at least 20  $\mu$m. Designed for diffraction-limited performance over the entire wavelength range, MIDIR will require an adaptive optics system; a cryogenically cooled system could offer optimal performance in the IR, and this is a critical aspect of the instrument design. We present here an overview of the project, including a discussion of MIDIR's science goals and a comparison with other infrared (IR) facilities planned in the next decade; top level requirements derived from these goals are outlined. We describe the optical and mechanical design work carried out in the context of a conceptual design study, and discuss some important issues to emerge from this work, related to the design, operation and calibration of the instrument. The impact of telescope optical design choices on the requirements for the MIDIR instrument is demonstrated.}

\bigskip
\section{MIDIR: The mid-IR instrument for the European Extremely Large Telescope}

\begin{figure}
\centering
\includegraphics[width=7 cm]{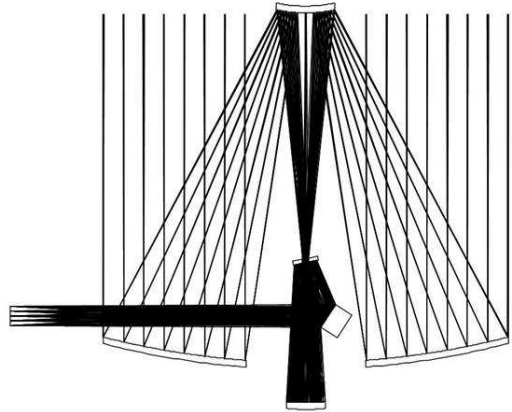}
\caption{Proposed optical design of the E-ELT. The design contains 5 mirrors, including a 42-m aspheric primary, a 6-m secondary and a 2.5-m flat adaptive mirror.}\label{fig:elt_optics}
\end{figure}

Since the mid-1990s the European astronomy community has been working towards a next-generation ground-based optical/infrared observatory, the European Extremely Large Telescope (E-ELT)\cite{gilmozzi07}. In 2006, a baseline reference design for the telescope was presented to the community, featuring a 42-m F/1 aspheric primary mirror in a 5-mirror optical design (see Figure~\ref{fig:elt_optics}). First light for the E-ELT is currently foreseen for 2017; an overview of the instrumentation program is given by Cunningham et al.\cite{colin06}.

MIDIR will be the mid-infrared instrument for the E-ELT. It
will cover the wavelength range from 3 to 20-24 $\mu$m and provide diffraction-limited imaging and spectroscopy at very high (R=50,000) resolution. Based on an earlier concept for the 100-m OWL telescope~\cite{towl_study}, a design study was completed in 2006\cite{midir_smallstudy} by an international consortium describing the science goals for a mid-{IR} {ELT} instrument and a conceptual design for MIDIR. An important aspect of the case for MIDIR is its complementarity to other infrared facilities that will be in operation in the next decade, most notably MIRI, the mid-{IR} imager/spectrograph for the James Webb Space Telescope (JWST). The study identified a number of issues that will form the subjects of a detailed trade-off study commencing in 2007/2008.

This paper gives an overview of the conceptual design of MIDIR and an outline of the critical issues. Section~\ref{science} presents highlights of the extensive science case for a mid-{IR} instrument on the {E-ELT}, from the study of conditions in the early solar system to follow-up observations of high-redshift gamma ray bursts. A brief comparison with other current and future IR facilities is also given. Based on these goals, Section~\ref{tlr} lists the top level requirements for MIDIR. Section~\ref{design} shows the conceptual optical and cryo-mechanical design for the instrument respectively, as well as the adaptive optics requirements for achieving diffraction-limited performance. Some critical issues are discussed in Section~\ref{issues}.

\bigskip
\section{Mid-IR astronomy in the ELT era}\label{science}

\begin{figure}[h]
\centering
\includegraphics[width=18cm]{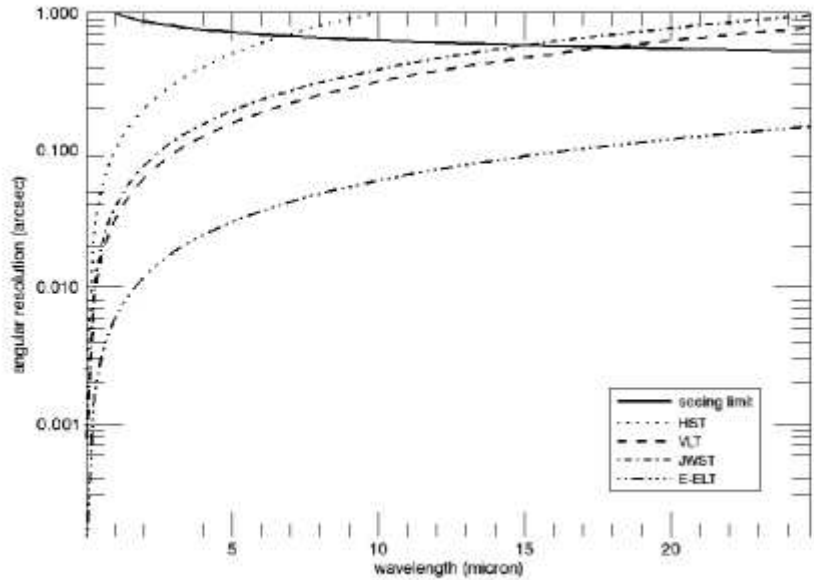}
\caption{Comparison of theoretical diffraction limit (given by $1.22 \lambda/D$) of the 42-m E-ELT with other facilities: the Very Large Telescope (VLT), Hubble Space Telescope (HST) and James Webb Space Telescope (JWST). Plotted for reference is the size of the seeing disk for $r_0$ of 0.8 arcsec at 500 nm.}\label{fig:difflt}
\end{figure}

The combination of a 42-m aperture and a good telescope site will open up new areas of IR astronomy to ground-based telescopes. While cooled space-based observatories have superior sensitivity in the mid-IR to faint extended sources, the larger aperture size of an ELT provides a higher angular resolution than is available in space. For example, at 10 $\mu m$ JWST has a diffraction limit of approximately 400 milli-arcseconds (mas); for the E-ELT this becomes 60 mas, a factor of 7 improvement. A plot of the diffraction limit of the E-ELT compared with that of a single VLT telescope, the Hubble Space Telescope (HST) and JWST is shown in Figure~\ref{fig:difflt}. For comparison, the size of the seeing disk for ground-based telescopes under average seeing conditions ($r_0$=0.8 arcsec @ 500 nm) is also plotted. In addition, no space-based mid-IR instruments provide the high spectral resolution foreseen for MIDIR. In terms of operation, a ground-based ELT instrument can offer quicker response times, and the telescope's large photon collecting area makes it more suitable for high time resolution and variability studies.

In order to exploit these advantages and provide complementarity with other contemporary IR facilities, most science cases for MIDIR involve the study of compact targets, rather than faint surface brightness objects. Exciting science cases for MIDIR focus on:

\begin{itemize}
\item highest angular resolution
\item very high spectral resolution
\item quick response time
\item time variability studies (milli-seconds to minutes)
\end{itemize}

This section will discuss in more detail some highlights from the MIDIR science case. However, the instrument's capabilities are by no means limited to these applications.

\bigskip
\subsection{Conditions in the early solar system}

Comets are thought to represent the most primordial bodies in the solar system accessible to Earth-based observations. The structure and composition of cometary ices, which are particularly sensitive to temperature and radiation processes, are key to our understanding of the formation and evolution of matter in the early solar system\cite{bockelee-morvan04}. Sensitive mid-IR spectroscopy can give an insight into the 'weathering' processes a comet has undergone since its formation. Using the high angular resolution of MIDIR, composition of comets' gaseous outflows can be studied within seconds of the material leaving the surface, ensuring that the composition has not yet altered once the molecules are released. Such studies have been carried out with the current generation of 8-10-m telescopes\cite{ootsubo07, harker05}, but have been limited to the brightest comets.

\bigskip
\subsection{Exoplanet detection}

To date more than 200 extrasolar planets have been detected, most of them only indirectly via the Doppler shift method. However, direct detections of the light emitted from the planets are needed to derive physical parameters such as temperature, chemical composition, and atmospheric structure and composition. To detect the radiation from the planet two general methods can be used:

\begin{enumerate}
\item spatially resolving the planet from the star, and
\item separating the radiation spectro-photometrically.
\end{enumerate}

While the detection of spatially resolved extrasolar planets around brown dwarfs has been reported in the literature\cite{chauvin05}, the search for planetary spectral signatures in spatially unresolved planet-star system seems very promising for MIDIR. The most effective approach is to select systems with planets that are known to transit their star, the number of which is set to increase significantly with upcoming space missions such as Kepler and Corot. Model spectra by Burrows et al.\cite{burrows03} of planets/brown dwarfs have illustrated the strong absorption features of molecular species such as methane and ammonia, which are not present in a pure stellar spectrum and, depending on the planet's position, will vary with time as the planet orbits the star . These features fall within the wavelength range covered by MIDIR.

\bigskip
\subsection{Circumstellar disks}

Circumstellar disks are a natural and important by-product of the star formation process. The material in these disks comprises the building blocks for planetary systems\cite{lissauer93, beckwith99}. Gaseous planets are thought to form within 1-10~MYr after the formation of the planet star; over time, the gas in the disk will dissipate and the remaining debris disk will host the formation of terrestrial bodies\cite{pollack96}. Observations of disks reveal a large source-to-source variation, suggesting a complex evolution from a gas-rich to gas-poor disk stage, and the transition phase is clearly pivotal in the planet formation process. Direct observations of the gas content, kinematics and composition are key to constraining the processes and timescales involved\cite{meyer06}.

\begin{figure}
\centering
\includegraphics[width=10cm]{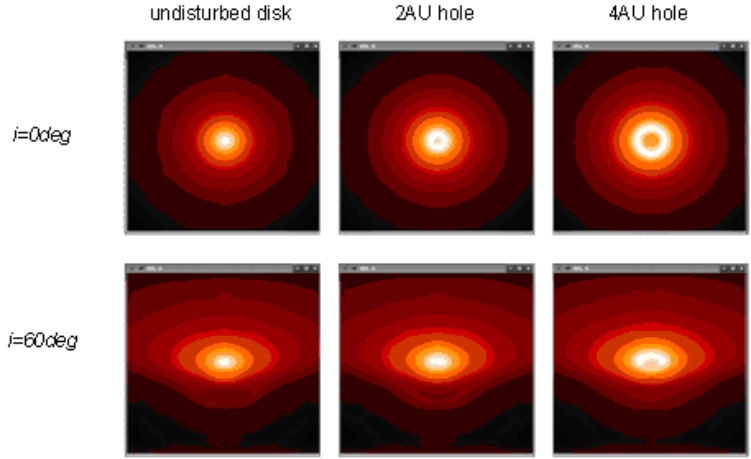}
\caption{Simulations of re-emitted light at 10 $\mu$m, assuming an unperturbed disk and a disk with a hole of radius 2 AU and 4 AU respectively, convolved with a 42m-ELT point spread function. The region corresponds to 60 $\times$ 60 AU. The star is assumed to be a Herbig Ae star (T=10,000 K, L=46L$_{\odot}$) at a distance of 60 pc; the disk parameters are those of the Butterfly star~\cite{wolf03} assuming a flared, Shakura-Sunyaev-type disk with a disk mass of 0.01 M$_{\odot}$ and an outer radius of 100 AU.  The dust grain size distribution and chemical composition are those of typical interstellar medium dust.}\label{fig:disks}
\end{figure}

Given the typical distances to nearby star-forming regions and the high angular resolution available on a 42-m ELT, MIDIR will have the ability to produce high-resolution images of the inner regions of circumstellar disks, which contain the signatures of planet formation. Simulations shown in Figure~\ref{fig:disks} show that gaps in the disks around intermediate mass stars at a distance of 60 pc can be detected with MIDIR on the E-ELT. The star is assumed to be a Herbig Ae star (T=10,000 K, L=46L$_{\odot}$) at a distance of 60 pc; the disk parameters are those of the Butterfly star~\cite{wolf03} assuming a flared, Shakura-Sunyaev-type disk with a disk mass of 0.01 M$_{\odot}$ and an outer radius of 100 AU.  The dust grain size distribution and chemical composition are those of typical interstellar medium dust.

High resolution spectroscopy will also allow the study of several diagnostics for studies of proto-planetary disks, such as the evolution of dust grains, the properties of organic molecules, polycyclic aromatic hydrocarbons (PAHs) and ices.

\bigskip
\subsection{The Galactic Centre}

The Galactic Centre region is of great interest not only as the centre of our galaxy but also as the environment of the closest (quiescent) super-massive black hole (SMBH). All constituents of the inner few parsecs (the black hole, surrounding star clusters, streamers of ionized gas, molecular dust ring and supernova remnants) have been extensively studied\cite{melia01}. Largely enshrouded by dust, the region is best studied at radio, sub-millimeter, IR, X-ray and gamma-ray wavelengths.

There are numerous topics of great scientific importance in this complex region, including stellar kinematics around the SMBH and its associated radio source SgrA*\cite{sesana07}, the accretion and emission mechanisms of the SMBH. Variability and flares have been observed at IR wavelengths\cite{genzel03}. With its high spatial resolution, MIDIR on the E-ELT will be able to detect SgrA* and measure broadband fluxes coming from the central region at the most critical wavelengths (L-M bands), and thus contribute greatly to our understanding of black hole accretion processes.

\bigskip
\subsection{Centres of nearby Active Galactic Nuclei}

The most extreme starbursts are found in the centres of mergers of gas-rich galaxies\cite{sanders88}. Ultra-luminous infrared galaxies (ULIRGS) are a manifestation of this phenomenon, and approach quasar-like luminosities, which are, however, almost entirely radiated at mid- and far-IR wavelengths. Although
locally rare, at high redshifts ULIRGs are responsible for a large fraction of the integrated sub-millimetre background, the overall star formation history of the Universe and the evolution of galaxies in general. MIDIR will not only penetrate the huge amounts of dust but also observe numerous diagnostic features, such as [Ne II], [S IV], [Ar III], [S III], PAH emission features, the molecular H$_2$ lines, and the broad 9.8 and 18 $\mu$m silicate features.

MIDIR will be able to resolve the circum-nuclear starburst and the molecular torus from the broad-line region for a large number of objects, and thus allow a comprehensive study of the large-scale gas transport and accretion processes to super-massive black holes. While MIDIR will not quite reach the angular resolution provided by the VLTI, its high sensitivity is necessary for a large sample of objects to disentangle physical from geometrical effects.
\bigskip
\subsection{High-redshift gamma ray bursts (GRBs)}

GRBs are of interest to probe their progenitors, their host galaxies and the ionization state and metal content of the intergalactic medium (IGM) at high redshifts.
Furthermore, at z$\sim$10 even the observed near-IR K-band corresponds to rest frame UV light which may be strongly affected by extinction. The fact that most GRBs at higher redshift are located in very low-dust host galaxies may be a selection effect, which may be relevant to the peculiar "dark bursts"\cite{rol05}. At 10 $\mu$m many GRBs have flux densities that can be easily imaged with MIDIR, and the brighter ones are within the capabilities of the spectrograph which may be used to determine redshifts via the Pa-$\alpha$ (1.87$\mu$m) line (3.0$<z<$6.2) or the Pa-$\beta$ (1.28 $\mu$m) line (4.8$<z<$9.5). Short response times are crucial and can be provided by the E-ELT, but not by JWST.

\bigskip
\subsection{Performance comparisons}\label{jwst}

The science case outlined above has shown that MIDIR has the potential to contribute immensely to our understanding of many phenomena in the Universe. Not only does it cover important niches in the parameter space, it will also be competitive with other contemporary space-based IR facilities in the next decade. Some of the main "competitors" for MIDIR are listed below.

\begin{description}
\item[JWST-MIRI.] JWST will be launched in 2013 into an orbit at the L2 Lagrange point, with a planned operational lifetime of 5 years (possibly extendable to 10 years). JWST will thus most likely have ended its operation when the E-ELT comes fully online. Although MIRI will offer unrivalled sensitivity and, the 42-m ELT aperture can provide a much higher spatial and spectral resolution, as well as faster response times.
\item[SAFIR.] The Single Aperture Far InfraRed Observatory, SAFIR, is a large cold (5K) space telescope with a proposed launch date of 2015-2020, covering the wavelength range of approx. 30 to 500 $\mu$m\cite{lester06}. Offering an imager and low-, medium- and high-resolution spectrometers, SAFIR could be a powerful complementary facility to MIDIR.
\item[WISE.] The Wide-field Infrared Survey Explorer (WISE), launch date 2009, will provide an all-sky survey from 3.5 to 23 $\mu$m up to 1000 times more sensitive than the IRAS survey\cite{mainzer05}. One of the main survey aims is to provide a data base for JWST and other mid-IR pointing facilities. Thus WISE is not competitive to JWST and/or MIDIR, neither in resolution nor sensitivity, but will provide a detailed study of selected astrophysical objects.
\end{description}

\begin{figure}[b]
\centering
\includegraphics[width=10cm]{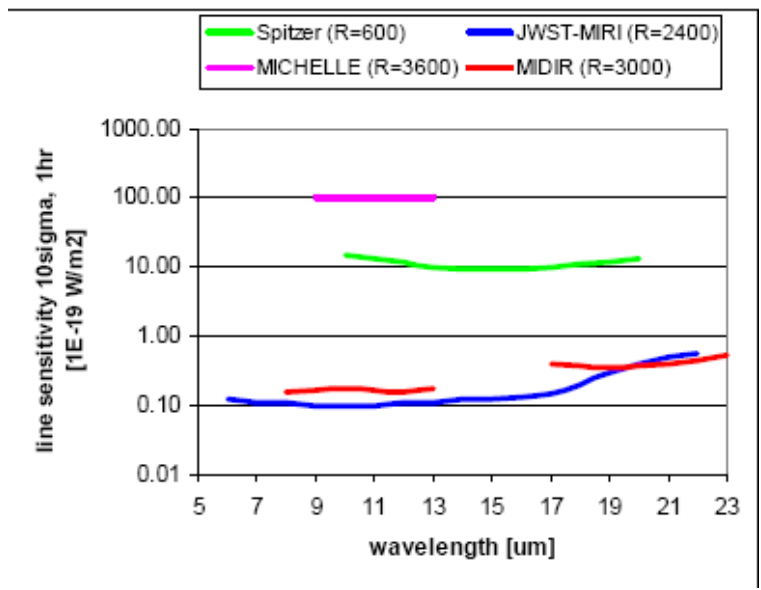}
\caption{Point source line sensitivities comparison between MIDIR on a 42-m E-ELT, JWST-MIRI, Gemini-MICHELLE and Spitzer-IRS for an unresolved line detected at 10-$\sigma$ in one hour.}\label{fig:sensitivities}
\end{figure}

Table~\ref{tab:tel_comparison} gives a summary of relevant instrument parameters for MIDIR compared with those of other mid-IR observatories. Data for IRAC are listed for comparison. Figure~\ref{fig:sensitivities} compares sensitivities between JWST, MIDIR and the current instruments Gemini-MICHELLE and Spitzer-IRS, showing MIDIR's excellent performance approaching the JWST-MIRI sensitivity but at higher angular resolution.

\begin{table}
\centering
\begin{tabular}{|l|p{2.5cm}|p{1.8cm}|p{2cm}|p{2cm}|p{1.5cm}|c|}
\hline
Project & Wavelength range ($\mu$m) & Telescope diameter (m) & Telescope temperature (K) & Diffraction limit @ 5 $\mu$m (mas) & FoV (arcmin) & Launch\\
\hline
JWST-MIRI\cite{wright04} & 5 - 28 & 6.5 & 50 & 159 & 2.3 $\times$ 2.3 & 2013\\
\hline
MIDIR & 3-24 & 42 & 290 & 30 & 1 $\times$ 1 & 2017\\
\hline
SAFIR\cite{lester06} & 30 - 500 & 10 & 5 & 620 @ 30 $\mu$m & 10 - 30 & 2020\\
\hline
Spitzer-IRAC\cite{fazio04} & 3.6 - 8 & 0.85 & 70 & 1213 & 5 $\times$ 5 & 2003\\
\hline
WISE\cite{mainzer05} & 3 - 25 & 0.40 & 15 & 2580 & 45 $\times$ 45 & 2009\\
\hline
\end{tabular}
\caption{Comparison of MIDIR's instrument parameters with those of other contemporary IR facilities.}\label{tab:tel_comparison}
\end{table}

\bigskip
\section{Top level requirements}\label{tlr}

For MIDIR to capitalise on the advantages offered by the large E-ELT aperture and high spatial and spectral resolution, the following top level requirements were identified:

\begin{itemize}
\item Wavelength coverage of 3 to 20 $\mu$m, with a goal to extend coverage to 27 $\mu$m, depending on the atmospheric transmission at the selected E-ELT site. This will cover the L, M, N and Q bands.
\item Four observing modes:
\subitem Camera for broad- and narrow-band imaging (approx. 30 filters) over a 20 $\times$ 20 arcsec FoV;
\subitem Low-resolution long-slit spectrograph (R$\sim$300);
\subitem Medium-resolution IFU spectrograph (R$\sim$3000); and
\subitem High-resolution IFU echelle spectrograph (R$\sim$50,000).
\item Diffraction limited performance over the entire wavelength range (Strehl $>$ 80\%).
\item Background (instrument + AO) $<<$ (sky + telescope) at all wavelengths and resolutions.
\item Several parallel observing modes.
\end{itemize}

The options of coronagraphy and polarimetry will be investigated as part of the Phase A study.

\begin{figure}
\centering
\includegraphics[width=12cm]{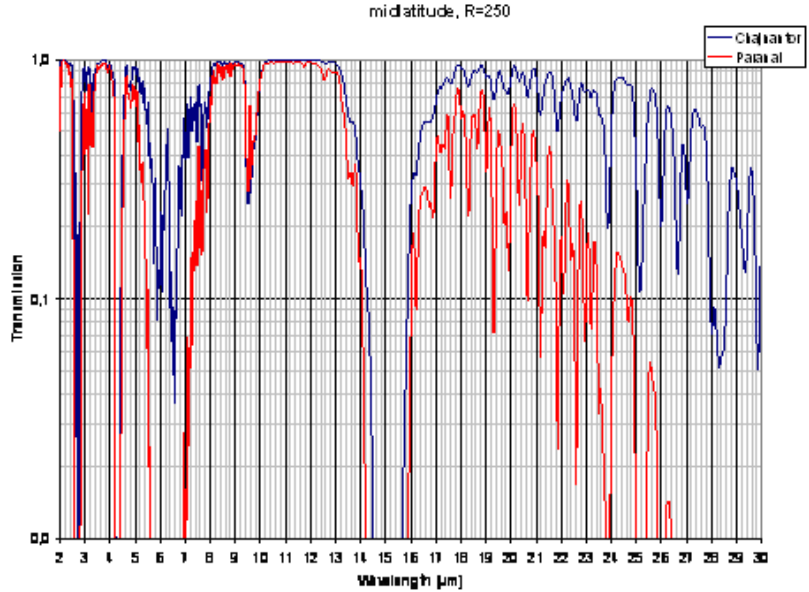}
\caption{Atmospheric transmission for two elevations, 2600 m (Paranal-like)  and 5100 m (Chajnantor-like).}\label{fig:transmission}
\end{figure}

The feasibility of the science case for MIDIR depends greatly on the atmospheric characteristics of the chosen E-ELT site. At M and Q bands the performance is mainly limited by the atmospheric transmission, while in the N-band the performance is largely given by the temperature of the atmosphere and telescope, as well as the transmission. Figure~\ref{fig:transmission} compares the typical atmospheric transmission at elevations of 2600 and 5100 m, showing the much-improved transmission at the higher site, particularly in the M and Q bands.

\begin{figure}
\centering
\includegraphics[width=11cm]{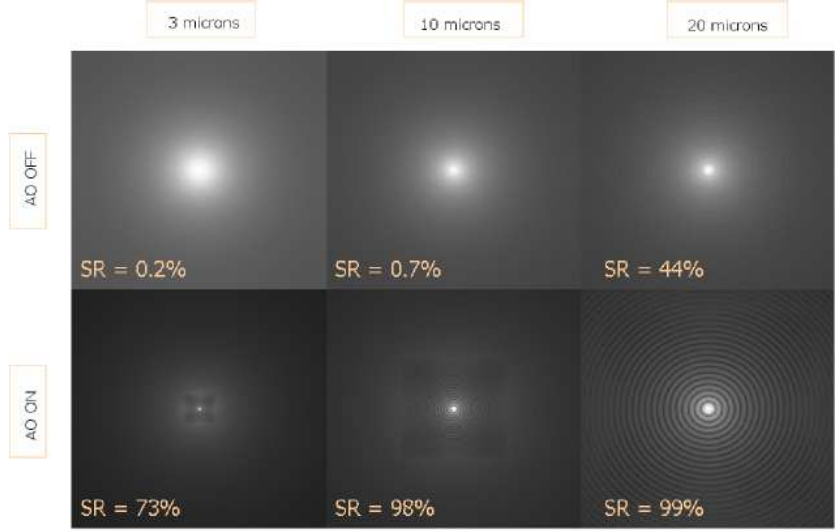}
\caption{Simulated adaptive optics correction on a 42-m E-ELT at mid-IR wavelengths under typical atmospheric and operational conditions: 0.7" seeing @ 500 nm, an outer scale of 60 m, subaperture size $=r_0$ @ 3 $\mu$m (1.24 m), wind speed of 10 m/s, wavefront sensor integration time of 2 ms, WFS time lag of 1 ms. A single natural guide star was used. The PSF is normalised to 1. (Image courtesy of L. Jolissaint)}\label{fig:ao_corr}
\end{figure}

The requirement for diffraction-limited performance at all wavelengths implies the need for a well-integrated adaptive optics (AO) system for MIDIR. While AO is not routinely employed at mid-IR wavelengths on today's 10-m-class telescopes, Figure~\ref{fig:difflt} illustrates the benefit of an AO system in this regime on a 42-m telescope: in the N-band the diffraction-limited resolution is an order of magnitude improvement on the seeing-limited value\footnote{The graph in Figure~\ref{fig:difflt} demonstrates that even observations with 10-m-class telescopes @ 10 $\mu$m can benefit from low-order turbulence correction; in practice improvement in resolution is achieved by implementation of the so-called `burst mode'\cite{doucet06}}. Using the analytic AO simulation code PAOLA \cite{paola}, the increase in Strehl ratio (SR) was modelled for a 42-m telescope using typical atmospheric and operational conditions for the E-ELT and a perfect AO system. Figure~\ref{fig:ao_corr} shows the resulting point source image at 3, 10 and 20 $\mu$m. Issues with the AO system design are discussed in Section~\ref{issues} and are described in more detail by Kendrew et al.\cite{kendrew_vancouver07}.

The performance of MIDIR is also affected by the location of the instrument at the telescope. The standard location for all instruments is the Nasmyth platform. However, the 5-mirror design (see Figure~\ref{fig:elt_optics}) offers the possibility of an intermediate focus position for MIDIR. The reduction in reflective surfaces would offer a reduction in telescope background but the location also poses significant technical and operational challenges. This will be further discussed in Section~\ref{issues}.
\bigskip
\section{Conceptual design for MIDIR}\label{design}

This section will give a brief overview of a conceptual optical and cryo-mechanical design for MIDIR.

\subsection{Optical Design}

The current MIDIR design concept favours a physically combined imager and spectrograph due to their commonality in operation (observing sequences, instrument control, detector readouts), background conditions (baffles and chopping) and telescope requirements. A novel modular design is envisaged, with the imager, low-resolution spectrometer, calibration source and possible AO system contained in a central cryostat. Three smaller cryostat units contain the optics for three medium- and high-resolution spectrograph channels, for the L/M, N and Q bands. Due to the very different requirements on the field of view of the imaging anf spectroscopy modes, the beams will be split immediately after the common pre-optics.

The optical design for the imager channel employs a purely reflective system, composed by a three-mirror anastigmat (TMA) collimator and two following cameras for thermal and mid-IR wavelength bands respectively. This solution provides the optimum resolution across the wavelength bands and allows for a compact design. Low resolution spectroscopy can be carried out in the imaging channel by inserting a grism into the collimated imager beam.

The system for spectrally filtering and spatially slicing the three spectral bands to the spectrometer channels employs similar technology to that developed for JWST-MIRI\cite{wells06}. The bands are sliced spectrally by a combination of two dichroics, each of which is designed to transmit long and reflect short wavelengths. An Integral Field Unit (IFU) slices the field of view spatially. A conceptual design for the spectrometer channels was also devised, optimised for the N-band requirements. The main drivers were the compactness of the design, and the possibility of using the same pre-optics and detectors for medium- and high-resolution arms of each channel. The optical layout of the spectrometers uses a TMA-based system for the high-resolution channel; the medium-resolution design employs technologies used for VISIR\cite{lagage00} and MIRI\cite{wells06}.

\begin{figure}
\centering
\includegraphics[width=16cm]{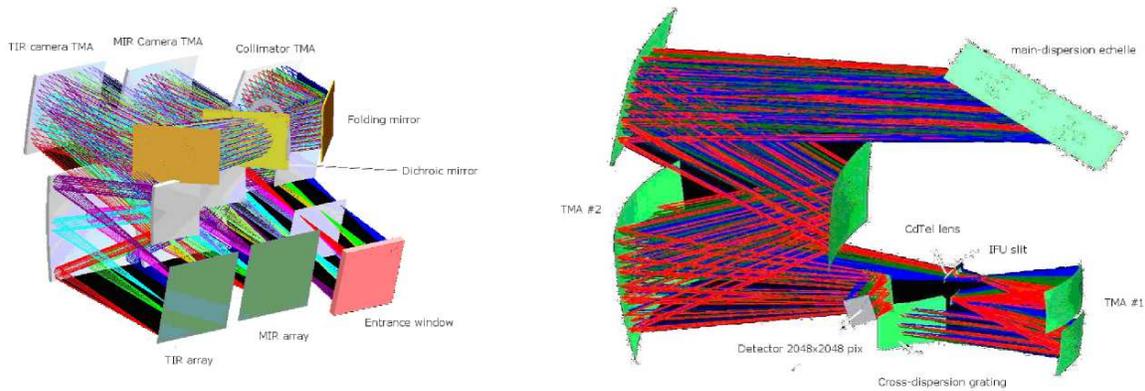}
\caption{3-D layouts of the imager (left) and one of the high-resolution spectrograph channels (right).}\label{fig:optics}
\end{figure}

Figure~\ref{fig:optics} illustrate the preliminary optical layout of the imager (left) and one of the high resolution spectrograph channels (right).

\bigskip
\subsection{Cryo-mechanical design}\label{design:mech}

\begin{figure}
\centering
\includegraphics[width=12cm]{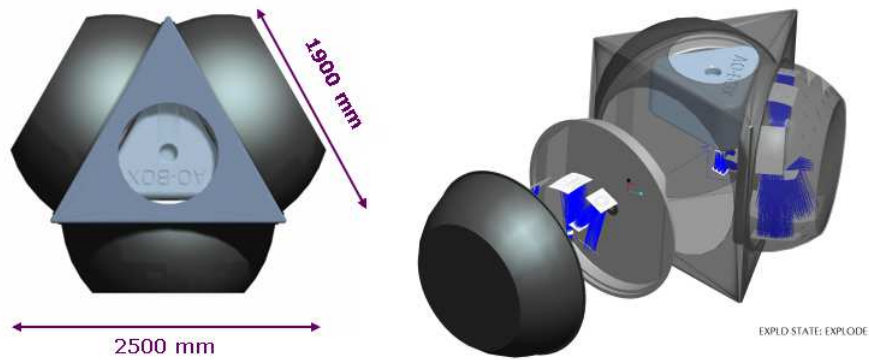}
\caption{Mechanical design of MIDIR, showing the top view of the unit (left) and an exploded view of one of the three smaller cryostat units (right.}\label{fig:mech}
\end{figure}

 The modularity of the optical design is reflected in the mechanical concept for MIDIR, with the main subsystems contained in coupled but separate small cryostats. The central triangular cryostat contains the pre-optics and imager optics (including the low-resolution spectroscopy optics); it could potentially hold an IR-optimised AO system should this be required. Its top flange is the mounting interface to the telescope and holds the cryostat entrance window. A calibration unit can be mounted on the warm side of this flange. The three smaller cryostats for the medium- and high-resolution spectrometer arms are mounted onto the three side faces of the central structure. These four cryostats have their own closed-cycle coolers but they are coupled to each other via the flanges of the central structure, and thus share a common vacuum. This is illustrated in Figure~\ref{fig:mech}.

Based on the temperature requirements for MIDIR, three cooling schemes are being considered:

\begin{itemize}
\item Pulse tube coolers (PTC)
\item Gifford McMahon coolers (GM)
\item Helium liquefiers
\end{itemize}

\noindent The exact choice of cooling system will be the subject of a trade-off analysis during MIDIR's Phase A study.

\bigskip
\section{Critical issues}\label{issues}

In the course of the conceptual design process, several technical and operational challenges emerged, which will be the subject of a detailed trade-off study commencing in 2007 as part of the Phase A study. Some critical aspects of the instrument design are summarised here.

\subsection{Location at the telescope}

The 5-mirror baseline reference design of the E-ELT (see Figure~\ref{fig:elt_optics}) provides access to 2 stable Nasmyth platforms, which are the default instrument locations at the telescope. For a mid-IR instrument, the presence of 5 warm mirrors in the optical path can cause a serious degradation in performance. The optical design of the telescope does, however, offer an intermediate focus position after only 2 reflective surfaces; placing MIDIR at this location improves the situation drastically (reducing the telescope emissivity by approximately 5\%~\cite{midir_sensitivity}), and offers the opportunity to incorporate an AO system optimised for mid-IR wavelengths in the instrument design, minimising any chromatic errors introduced from using a NIR-optimised AO system.

Placing MIDIR at the intermediate focus would, however, introduced significant added complexity. First, experience with full cryogenic AO systems is non-existent and this would entail a high technical risk. In terms of operation, MIDIR would have to interface directly with the primary mirror co-phasing and correct for segment misalignment at the instrument level. Access to the instrument at the intermediate focus would also be more complicated, making routine maintenance challenging. The feasibility of using the intermediate focus position for MIDIR must therefore be studied very carefully by both the instrument and telescope design teams.
\bigskip
\subsection{Adaptive optics and the mid-IR atmosphere}

The AO requirements for MIDIR will also form the subject of a detailed trade-off study and are closely tied to the chosen location for MIDIR at the telescope, as discussed above. The inclusion of a mid-IR-optimised system into the central cryostat is likely to pose significant technical challenges as no experience is available with cryogenic AO systems. However, if MIDIR is located on the Nasmyth platform it will make use of the telescope's AO system, including the deformable secondary mirror (M4), which are optimised for NIR wavelengths. This will require careful design of the instrument to minimise the chromatic errors introduced by the difference between sensing and observing wavelengths.

Specification of AO requirements for MIDIR is further complicated by our lack of understanding of atmospheric turbulence at infrared wavelengths. At optical/NIR wavelengths atmospheric turbulence is caused by temperature variations; this process is relatively well understood and described analytically by the Kolmogorov and von Karman models\cite{hardy}. Infrared radiation is affected by variations in composition, as well as temperature fluctuations. Non-uniform mixing of water vapour in the atmosphere gives rise to spatial and temporal variations in transmission and emissivity of the atmosphere at IR wavelengths. The horizontal, vertical and temporal profiles of water vapour fluctuations on the scales of an ELT aperture, and how they affect IR image quality, are poorly understood.

Water vapour fluctuation measurements are available mainly from interferometric observations with VLTI or the Keck Interferometer\cite{koresko06, colavita04}; more recently VISIR was also used in conjunction with optical wavefront sensor data to collect atmospheric data\cite{tokovinin07}. Further characterisation of this phenomenon is needed to improve our understanding of its effects on E-ELT images in the IR.

\bigskip
\subsection{Chopping}

Astronomical observations in the thermal IR  with uncompromised sensitivity depend on signal modulation, also known as chopping. The method used by 10-m-class telescopes, i.e. secondary mirror chopping, is no longer feasible at ELTs. Several alternative techniques are available; their feasibility and suitability for MIDIR will be examined in detail during the Phase A study for MIDIR.

\begin{description}
\item[Focal plane chopping.] Chopping at the focal plane could be considered if carried at the detector. If the required chop throw can be provided mechanically by the detector this may be an attractive solution for compact objects with extents less than the chop throw.
\item[Pupil plane chopping.] In the case where MIDIR has its own AO system, pupil plane chopping may be possible using the tip-tilt mirror. The associated wavefront error introduced would have to be compensated by the AO system. However, this does introduce added complexity with respect to the interaction with the active optics system.
\item[Dicke switching.] This technique, used in radio astronomy, has been applied successfully in the past in IR astronomy\cite{deming86}. It is most likely the only method of exploiting a field of 20-30 arcsec without compromising the spatial information resulting from chopping.
\end{description}
\bigskip
\subsection{Operations}

This section is not applicable to MIDIR alone, but will apply for all new ELT instrumentation. Due to the expensive and valuable ELT-time, the required amount of telescope time for each observation should be minimised and used for scientific sky observations as much as possible. This requires that each instrument should calibrate and prepare itself, to the extent that only the effects of telescope and atmosphere require on-sky characterisation. This requires much more attention to calibration and instrument operation during the design and development of instruments, in order to facilitate instrument operation regardless of its design complexity.
\bigskip
\section{Conclusions}\label{conlusions}

The science case outlined in this paper has demonstrated that a mid-IR instrument on the E-ELT will open up new areas of research to ground-based IR astronomy; indeed, most science goals can only or better be done by MIDIR than with any other ground- or space-based instrument.

The conceptual design of MIDIR represents the first important step toward a real mid-IR ELT instrument. Although the instrument does not depend on fundamentally new technologies, certain technologies need to be further developed, and additional design and operational aspects need to be investigated. These issues will form the subject of a detailed trade-off study as part of Phase A for MIDIR, which is likely to commence in 2007/2008.

\bibliography{midir_bib}
\bibliographystyle{spiebib}

\end{document}